This is authors' copy of the manuscript accepted for journal Computer Supported Cooperative Work (CSCW). Please, refer to the published article at doi.org/10.1007/s10606-020-09377-x for further information.

Cite as:

Dolata, M., Schenk, B., Fuhrer, J., Marti, A., and Schwabe, G. When the System Does Not Fit: Coping Strategies of Employment Consultants. Comput Supported Coop Work 29, 657–696 (2020). https://doi.org/10.1007/s10606-020-09377-x

# When the System does not Fit: Coping Strategies of Employment Consultants

Mateusz Dolata[1✉], Birgit Schenk[2],
Jara Fuhrer[3], Alina Marti[3], Gerhard Schwabe[3]

[1✉] Department of Informatics, University of Zurich
 *dolata@ifi.uzh.ch*

[2] University of Applied Sciences - Public Administration and Finance Ludwigsburg
 *birgit.schenk@hs-ludwigsburg.de*

[3] Department of Informatics, University of Zurich
 *jara.fuhrer@uzh.ch, alina.marti@uzh.ch, schwabe@ifi.uzh.ch*

**Abstract.** Case and knowledge management systems are spread at the frontline across public agencies. However, such systems are dedicated for the collaboration within the agency rather than for the face-to-face interaction with the clients. If used as a collaborative resource at the frontline, case and knowledge management systems might disturb the service provision by displaying unfiltered internal information, disclosing private data of other clients, or revealing the limits of frontline employees' competence (if they cannot explain something) or their authority (if they cannot override something). Observation in the German Public Employment Agency shows that employment consultants make use of various coping strategies during face-to-face consultations to extend existing boundaries set by the case and knowledge management systems and by the rules considering their usage. The analysis of these coping strategies unveils the forces that shape the conduct of employment consultants during their contacts with clients: the consultants' own understanding of work, the actual and the perceived needs of the clients, and the political mission as well as the internal rules of the employment agency. The findings form a twofold contribution: First, they contribute to the discourse on work in employment agencies by illustrating how the complexities of social welfare apparatus demonstrate themselves in singular behavioural patterns. Second, they contribute to the discourse on screen-level bureaucracy by depicting the consultants as active and conscious mediators rather than passive interfaces between the system and the client.

# 1  Introduction

The standard frontline setting in public agencies involves a computer running a case and knowledge management system and a monitor turned towards the frontline employee. Everyone knows the situation, in which one sits in front of a clerk using the computer and waits till they finish their "computer work". The complexity of the tasks conducted by the clerk in that moment remains hidden from the client unless the former decides to share aspects of their work by saying what they are doing or by showing it on the screen. The situation might make the client think that the computer rules the interaction between them and the frontline employee as suggested by previous research on consultation encounters (Pearce et al., 2012; Kilic et al., 2015; Kilic et al., 2017). There are two opposite claims on how technology impacts the frontline work of public agencies: it can curtail or empower the personnel. A curtailment situation exists if the frontline employee does not take any decision at their discretion – they act as if they were a human interface between the client/citizen and the computer (Snellen, 2002). But, computers also get appropriated in a way that makes the frontline employees process more work in less time (Jorna and Wagenaar, 2007) or allows them to offer more comprehensive and client-oriented services (Giesbrecht et al., 2014; Giesbrecht et al., 2016a; Giesbrecht et al., 2016b). The discussion on the impact of IT on frontline interaction in public agencies emerged decades ago (Lipsky, 2010) and, with the growing role of IT in provision of public services, still remains highly relevant.

Accordingly, there have been efforts to explain the effects of IT on the frontline interaction. Some CSCW and IS literature points to the design of the system (Pearce et al., 2008; Giesbrecht et al., 2013; Kilic et al., 2015) or the training of frontline employees (Giesbrecht et al., 2015; Giesbrecht et al., 2016a) as explanation for the IT's impact. Sociology points out that the interplay between the IT and the inherent culture of the institution favours either curtailment or empowerment effects (Bovens and Zouridis, 2002; Hansen et al., 2018). Finally, some researchers observe a general shift in the context of work (e.g., expectations from society), which in turn impacts the actual behaviour of the frontline employees towards the computer (Hupe and Buffat, 2014). Those general explanations often lack acknowledgement of the frontline employees as independent, autonomous agents who deliberately choose what to do and when. CSCW research has a long tradition in emphasizing the agency of workers and positioning them as subjects in complex work settings, as explicated in the appropriation research or the acknowledgement of individual values and practices in shaping the work environment (Nardi and O'Day, 1999). Taking this perspective, we assume that the frontline employees sometimes may deliberately let the computer and rules curtail their behaviour while otherwise they may take additional effort to overcome the curtailment. This study explores various coping strategies and the reasoning behind them. The way frontline employees reflect on and rationalize their use of technology reaffirms their status as active and conscious subjects in this system.

Understanding the motives for varying behavioural patterns of frontline employees is of high relevance to the public service. Many countries proclaim the change towards citizen-centric services and discuss ways to improve the overall service quality. Those efforts are accompanied by slogans and campaigns, as well as hierarchical changes or structural reorganization (less or more centralization, changes in processes) (Lipsky, 2010). However, the ones who are most impacted by the change are often left without the right tools, without adequate training, and, sometimes, without an updated regulatory basis. As a consequence, frontline employees face an increasing variety of challenges – in particular, they experience a conflict between the incompatible visions specified in the government roadmaps, the actual regulations implemented in the agencies, the technology provided to support their work and the expectations of the citizens. Whenever dealing with a citizen, the frontline employee may choose to follow a path reflecting one or the other vision. Without understanding and addressing the frontline employees' values, claims and slogans about citizen-centric public services are rather likely to remain wishful thinking than to become reality.

This study focuses on the work of employment consultants at the German public employment agency (Agentur für Arbeit, short *AA*). The primary mission of AA is to implement the vision of the German federal government considering the labour market by providing placement services, employment promotion services and by distributing the unemployment benefits. AA does not offer integrated welfare services (separated by law): low-income individuals or families, as well as part-time jobbers or self-employed do not receive support from AA unless they are affected by unemployment. The study focuses on the frontline work of employment consultants. The consultants are clerks who received an additional training concerning job placement, career planning as well as labour market issues. Their primary task at work (taking up to 80% of their work pensum) is consultations with clients, i.e. unemployed or job-seeking residents. The clients are individuals who have lost their last job, or intend to quit their position, or are entering the German labour market. Some clients attend to the consultations voluntarily (e.g., to learn about labour market situation before they decide to look for a higher-paid job), but the majority of clients are recipients of the unemployment allowance. For them, the consultations are an obligatory part of the unemployment reduction program: if they do not attend the consultations or they go against the agreed work integration measures, they might face punishment in form of reduced unemployment benefits. Even though employment consultants are required to inform other departments of the AA about such facts, they are not directly involved in decisions concerning material or monetary allowances. Monetary support depends on the last wage, the date of unemployment insurance, the duration of unemployment and the willingness to take up a new job (Schmitz, 2013).

The counselling services are presented as essential for getting the clients back to the labour market. Through participation in a service, clients receive help and support in their individual efforts to find a position and to terminate the unemployment (AA, 2018a). Through regular consultation with clients, the AA controls whether they actually undertake the necessary actions and if they can therefore fulfil their regulatory duty of support and supervision (AA, 2016; Baldauf, 2016). However, given the compulsory character, clients may experience the service as a burden instead of support (Shore and Tosun, 2017). This puts additional pressure on the quality of the consultation. If the consultation feels like a screen-driven data collection, clients are unlikely to see its actual value and to implement the consultant's suggestions. However, if they experience an empathetic and individualized service, they are likely to appreciate the continuous collaboration with a consultant (Böhringer, 2015).

Accordingly, research and practice continuously call for improvement of frontline experience in employment agencies (Böhringer, 2015; Roman, 2015). The proposed improvements focus primarily on the organizational processes (Roman, 2015), the self-services (Bovens and Zouridis, 2002) and the counselling techniques (Bezanson, 2004). Even though CSCW research has proposed the redesign of information technology (IT) as a way to enhance work and collaboration in welfare agencies (Boulus-Rødje, 2018), research in public management shows that IT is primarily used as a way to document and store the client's profile, search for job offerings and match openings with profiles (Holt and Huber, 1969; Sampson and Reardon, 1998; Buffat, 2015). AA's official documents and directives clearly state that its digital infrastructure was built to support documentation and control (Kupka and Osiander, 2017; AA, 2018b), similar to other employment agencies as depicted in public administration literature (Bovens and Zouridis, 2002; Van Berkel et al., 2017). Consequently, the consultants' lack dedicated tools designed for the collaborative setting of a consultation and are obliged to collect data with the systems provided. As a consequence, they appropriate the IT infrastructure provided in spite of its inconsistencies with their individual aspirations, local processes of their AA branch, or needs of the involved stakeholders. This study depicts the behavioural patterns involved in balancing the inconsistencies and which aspects advisors do have in mind when being involved.

We ask the following questions:

*RQ1. How do employment consultants cope with inconsistencies between IT structures and other demands during an unemployment consultation?*

*RQ2. What are the reasons behind these specific coping behavioural strategies of employment consultants?*

This study focuses on the observable, material consultation behaviour that goes against the scenarios prescribed by AA for the use of the system and that contradicts or bypasses the AA's general directives. These behavioural patterns are just a fraction of job consultants' work – we do not aim at a comprehensible representation of the complexity of decisions and actions taken by the consultants, especially outside of their consultations. We refer the reader to other studies that cover the job consultants' work in a more holistic manner (Roman, 2015; Van Berkel et al., 2017; Boulus-Rødje, 2018). However, the examination of the coping behavioural patterns provides a focused lens for the identification of the various considerations, reasons, and rationales that drive the consultants to behave one way or the other. Even though the behaviour might seem insignificant at first (e.g., turning the monitor), discussing those episodes with the consultants revealed their core values (e.g., transparency, understanding, empathy, esteem and respect towards the client). We claim that the awareness of these values is necessary to develop more adequate systems for supporting consultants at their work.

This work adds to the line of research on the implicit motives of consultants' work in various settings (Kilic et al., 2015; Kilic et al., 2016; Dolata and Schwabe, 2017b; Kilic et al., 2017; Dolata and Schwabe, 2018; Dolata et al., 2019). It also extends the discourse on the support of welfare workers in CSCW (Boulus-Rødje, 2018; Holten Møller et al., 2019; Karusala et al., 2019). Finally, it provides an argument in the ongoing discussion of the curtailment vs. the empowerment aspects of IT in frontline interactions at public agencies (Landsbergen, 2004; Buffat, 2015). Thanks to this study, employment agencies can obtain insight into the considerations of their employees and adjust their training and implementation policies. The consultants can benefit by learning how their colleagues employ the available resources and by receiving more appropriate support tools in a longer perspective. The clients can also hope for better consultation services with more adequate tools and more satisfied consultants.

## 2 Related work

### 2.1 Technology in consultations at public agencies

The wave of digitalization reached the public agencies in the 1990-ies. While it first changed management and back office structure, it soon also affected the street-level bureaucrats (Bovens and Zouridis, 2002; Lipsky, 2010), i.e., the clerks and civil servants who have contact with the general public in welfare services or rule enforcement. Whereas they previously had power to take decisions on their own or implement their own routines, through digitalization their freedoms were diminished. The need to document the initial situation and the actions taken allowed for control and standardization. This reduced the potential for discrimination and differences between singular caseworkers (Bovens and Zouridis, 2002), but also limited decisions taken in favour of the clients (Hansen et al., 2018). Soon after this, the systems took over some of the decision making, as the legal rules and organizational policy got implemented directly in the software (Hupe and Buffat, 2014; Hansen et al., 2018). Most public agencies, including the AA, employ case and knowledge management systems equipped with hard-coded decision trees and rules.

As a result, the street-level bureaucracy turned into the *screen-level* bureaucracy. In the screen-level bureaucracy, the frontline employees are constantly connected with the organization through the computer and it is the *screen* which drives their action (Bovens and Zouridis, 2002). The screen suggests which questions in which order to ask and what the possible answers are (Hansen et al., 2018). In the end, the system generates options on how to proceed. Given the importance of the computer screen, it is prominently placed on the table and turned towards the frontline employee (cf. Figure 1). In the past, researchers predicted this situation to be an intermediate stop on the way towards a *system-level* bureaucracy (Bovens and Zouridis, 2002; Hansen et al., 2018), where the clients directly interact with the system. However, the shift towards the system-level bureaucracy has not happened yet.

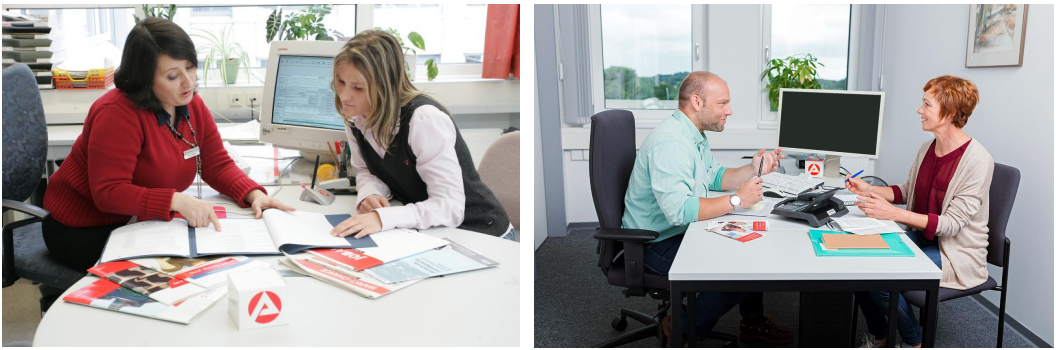

Figure 1. Illustrations of the employment consultations from the official materials of the German public employment agency (Agentur für Arbeit). On the left, a picture from around 2000; on the right a picture from 2017.

As a consequence, screen-level bureaucracy persists and thrives in numerous public agencies around the world. The face-to-face consultations remain the main contact channel for welfare services and employment services even though some agencies offer alternative interaction channels (Scott, 2004; OECD, 2015; Hansen et al., 2018). Many clients still stick to the seemingly more cumbersome face-to-face service and some agencies, like AA in Germany, see a regular face-to-face contact as necessary. Moreover, the web services are frequently used, not as a replacement, but as a complementary contact channel for interaction with public employment agencies (Kirk, 2000; OECD, 2015; Hansen et al., 2018). The preferences of the involved actors seem to contradict the claim that the screen-level bureaucracy was just an intermediate step: the face-to-face encounter involving the use of a desktop computer forms the current state-of-the-art in numerous domains (Great-batch et al., 1993; Greatbatch et al., 1995; Pearce et al., 2008; Pearce et al., 2012; Kilic et al., 2016; Kilic et al., 2017; Boulus-Rødje, 2018; Mørck et al., 2018). In other words, the argument about the future rise of pure system-level bureaucracy misses some essential advantages that keep face-to-face encounters alive, as suggested by the CSCW account emphasizing the role of workers as key elements in their environment (Greenberg, 1991; Nardi, 1996; Nardi and O'Day, 1999; Wulf et al., 2011).

The literature suggests that the clients prefer the face-to-face contact rather than self-services for several reasons. Many clients claim that the self-service algorithms and wizards do not fit their particular situation (Hansen et al., 2018). Sometimes, clients find the user interface too complicated (Hansen et al., 2018). Furthermore, analyses point to the distribution of preferences along the line of digital divide e.g. older and less educated people choose a personal contact since they have less experience with, or less access to self-services in general (Bratteteig and Verne, 2012; OECD, 2015; Shore and Tosun, 2017; Hansen et al., 2018). On the other hand because self-services are rarely offered in an attractive form (e.g., mobile applications) younger clients might avoid them, too (OECD, 2015). We see these explanations as valid but they seem to lack the consideration of the consultant as the actual differentiating aspect between today's face-to-face consultations and self-service envisioned in system-level bureaucracy (Bovens and Zouridis, 2002). Some recent studies point out that the surplus of talking to human consultants lies in their ability to understand the client's situation in an holistic, empathetic way and to ask the adequate questions (Dolata et al., 2016; Giesbrecht et al., 2016a; Dolata and Schwabe, 2017a). Both is problematic to computers, which cannot really *experience* how it is to be a human in a specific context and cannot make any inferences based on this (McAfee and Brynjolfsson, 2017). This study tries to describe the additional value generated by the consultants when they interact with their clients by pointing out coping strategies which they employ to moderate between IT and other demands.

In parallel, the literature discussed the nature of technology's impact on frontline interactions. Two accounts emerged: The *curtailment* argument claims that the discretion of frontline employees is limited more and more by technology thus eventually leading to full automation (Bovens and Zouridis, 2002; Snellen, 2002; Lipsky, 2010; Buffat, 2015). The *enablement* argument claims that IT provides valuable resources for employees' decision making and generates an illusion of control for managers, who reduce the direct controlling of their subordinates (Bovens and Zouridis, 2002; Jorna and Wagenaar, 2007; Buffat, 2015). The arguments were tested and supported in a range of organizational, longitudinal and theoretical studies (Landsbergen, 2004; Buffat, 2015; Hansen et al., 2018), but the actual impact on the interaction in frontline remains under-explored despite singular studies (Weller, 2006; Toerien et al., 2015; Boulus-Rødje, 2018). Past CSCW research on the impact of IT in consultations and advisory services beyond the context of employment agencies suggests that particular features of such a system (ranging from its format factor, over visualization of the content, up to the content itself) can lead to the emergence of favourable practices (e.g., Fischer et al., 2017; Bossen et al., 2019; Dolata et al., 2019) as well as practices that contradict the idea of the service (e.g., Kilic et al., 2015; Kilic et al., 2016; Mørck et al., 2018; Pine et al., 2018). Thus, we claim that the discussion about the appropriation of the case and knowledge management systems in public administration requires a consideration of particular features and the ways they get appropriated, as opposite to general accounts, which treat the technology as a black box. This study aims at exploring the curtailing and enabling elements of technology in practice and how those elements activate specific considerations and value-dilemmas in the employment consultants.

This perspective aligns with the CSCW interest in practice-oriented research and with the discourse on the appropriation of technologies at work. In particular, we focus on practices in their complex nature ranging from determination of the nature of the situation, through the selection of techniques, situated action in the context of work, up to the expected outcome of the employed coping behavioural patterns (Schmidt, 2014). It is important to note that complex systems and tools are not appropriated as a whole but rather based on the features that become salient in a given situation (Norman, 1988; Weilenmann and Lymer, 2014). Consequently, the study of how systems get appropriated in practice, e.g., during a consultation at an employment agency, requires a reflection on the features offered by the system and how those features interact with structures provided by the institutional context, values systems, regulations, rules, the material environment or even the ongoing political and societal discourses (Dourish, 2001; Orlikowski and Barley, 2001; Scollon, 2001; Orlikowski, 2008). These frame our approach towards work practices at employment agencies and emphasizes our interest in understanding the motives behind the observed material behaviour and how far those behavioural patterns depend on the available structures.

The systems used in employment agencies are normally case and knowledge management systems (Van Berkel et al., 2017). In particular, AA introduced a system to store information about the clients, to document their efforts to find a position, to search for appropriate positions, to register any agreements made between the consultant and the client, as well as to record other actions of the consultants in respect of a client (AA, 2013a; Toerien et al., 2015; AA, 2017; AA, 2018b). This information can be accessed by the consultants, other frontline employees, as well as back office employees responsible for allowances (AA, 2013a; AA, 2013b; AA, 2018b). The clients can access the most relevant information easily through the self-service system and, thanks to the General Data Protection Regulation, has access to all data upon request. Data collection and processing features are the main functions common for systems used by employment agencies in the developed countries of Western Europe and North America (OECD, 2015; Van Berkel et al., 2017; Hansen et al., 2018). The consultant can access the data through a user interface reflecting the database structure by using a software application deployed on a desktop computer. The typical control elements used in the interface involve text fields, buttons, check boxes, lists and links (AA, 2013a; AA, 2018b). Above all, the systems used in public employment agencies implement the information and knowledge processing needs of the agency and offer features to complete its regulatory duties (Bratteteig and Verne, 2012).

## 2.2 The role of consultants in employment services

Public employment agencies offer similar services despite geographical variance. Their structure, legal form and responsibilities differ across countries or regions (OECD, 2015). In some countries one central or some regional agencies take over legally assigned duties (e.g., Germany, Switzerland, France) while in other countries (e.g., Australia) employment agencies are private businesses (Marston, 2006; Campbell et al., 2013; OECD, 2015; Kupka and Osiander, 2017). Despite the structural differences, those agencies provide services concerned with job placement, labour market information and labour migration. In many countries they also distribute jobseeker allowances and other social benefits as well as enforcing governmental policies on the labour market (OECD, 2015; Van Berkel et al., 2017). In Germany the federal employment agency, with its ten regional directorates and 156 local agencies, administers the allowances and offers labour market services (AA, 2018a). The AA is not an integrated welfare office but an employment agency. The clients are people who have lost their job, quit their job, or simply want to change their job in the foreseeable future. Each local agency offers consultations for unemployed individuals at one or several branches. AA hires over 90,000 employees but only a fraction of them are job consultants (OECD, 2015; AA, 2018c). Each day, they conduct 14,000 full job placement consultations (AA, 2018c) and each consultant has approx. 5 to 7 consultation slots per day. This makes the counselling offer easy to access for the clients but also shows the impact of AA on the general public.

The main focus of consultations at AA is job placement. During an individual conversation, a consultant (also referred to as a *placement officer*) and the client devise an action plan and define job search activities and targets (OECD, 2015). The AA propagates a four-phase model (4PM) for counselling activities designed running across four topics: profiling, goal, strategy, and implementation (Schmitz, 2013; Kupka and Osiander, 2017). During *profiling*, the participants identify client's strengths and potentials and specify contextual data like the family. During *goal selection* they agree on a job profile as the most-promising target for the job placement activities. During *strategy selection* the participants develop a plan as to how to reach the identified target given the client's profile; a strategy might involve a vocational training or simply applying for an open position. The federal agency offers a list of over 35 applicable strategies but encourages local agencies to extend and adapt them to their particular situation. Finally, the *implementation* starts after the parties have settled on an integration agreement which includes the selected strategies and lasts for the following six months. The official guidance directs the consultant to structure and determine the agreement together with the clients, print it out and hand it to the clients during the consultation (Schmitz, 2013). During the first consultation, the focus should be on the profiling, as well as the selection, of goals and strategy; the strategies should be formulated as concretely as possible, thus including particular training opportunities or particular positions to apply for (Schmitz, 2013). The subsequent consultations support implementations by checking progress, adjusting plans, etc. The instructions for the consultants request them to document each step while using specific functions of the system (AA, 2013b; Schmitz, 2013). On the one hand, the description of the process underlines the key role of the first consultation and how important it is to get to know the clients but, on the other hand, it makes clear that the documentation and formal agreements are a critical element.

While the AA sees the role of a consultant primarily in terms of the task, he/she conducts during a consultation, their own understanding of their role is more nuanced. According to a survey among 178 AA consultants, they see themselves primarily as service providers and advisors (64%) as opposite to labour market broker (8%), social worker (12%) or bureaucrats (13%) (Osiander and Steinke, 2011; Kupka and Osiander, 2017). Further, they identify a *sustainable* reintegration into the labour market as their highest job objective (58%) as opposite to a *quick* reintegration (40%) or stabilization of the client's situation (2%) (Boockmann et al., 2013; Kupka and Osiander, 2017). This already points to a potential conflict between the guidelines and the consultant's own vision of the job. The official guidance calls for prioritization of the most promising goals and strategies with the client to shorten the duration of the unemployment, as well as the usage of binding and scheduled agreements. However, consultants seem to value sustainability higher (Schmitz, 2013; AA, 2016; AA, 2017; Kupka and Osiander, 2017). It remains open how the case and knowledge management system fits into this prioritization issue.

However, consultants follow the official guidelines in terms of the activities they undertake. They admit in the referenced survey that, most often, they deal with the following issues in their consultations: labour-market perspectives, integration agreement, job offers and vacancies, data collection or verification and application files (Boockmann et al., 2013; Kupka and Osiander, 2017). This is supported by another survey study, which also shows that the variance between particular local employment agencies is rather low if it comes to job understanding and conducted tasks (Roman, 2015). Whereas the questionnaire research provides essential insight into employment agencies in Germany, it relies strongly on self-reports, as opposed to Nordic countries or Australia where observational data from welfare centres has been published (Marston, 2006; Boulus-Rødje, 2018; Hansen et al., 2018). Because of the differences between the countries (e.g., no institutional distinction between basic income and jobseeker allowance in Norway or Denmark and private employment agencies in Australia), insights are hard to transfer. Additionally, survey research pays little attention to the role of IT, which may interfere with the private goals of consultants and clients and with particular situations during the consultation. Consequently, this study was launched to improve the understanding the connections between the employed system and the role of consultants in the AA.

## 3   Methodology

Given that the literature misses accounts of consultants' work with IT in the consultation encounters, we launched an ethno-methodologically informed fieldwork study in the real work environment of the consultants. In our study, we followed the standards of workplace studies (Heath and Luff, 2000; Luff et al., 2000) widely employed in the CSCW community and beyond, to analyse and describe situated work practices and structures that underlie them (Orlikowski, 2008; Schmidt, 2011; Wulf et al., 2011; Suchman, 2016; Dolata and Schwabe, 2018). Accordingly, two observers joined two local employment agencies in southern Germany in April 2018. Each observer spent overall approx. 15 working days in their assigned employment agency. Apart from that, each observer spent one additional day at the other agency to identify differences and similarities. The extended period of time dedicated to observation enabled the shadowing of consultants at work, the experience of the institutional context, as well as an informal interaction with the consultants. The key part the research stay was the visitation of consultations accompanied by interviews with the consultants and, if it was possible, with the clients. Above all, the study was designed to provide a broad and general understanding of employment consultants' work with special focus on their conduct in the consultation services.

During observation, the observers focused on the behavioural patterns of the consultants as material manifestation of work practices. Thus, it was important to see the activities happening in consultations and how they are triggered by artefacts in general and the software in particular. Consequently, the on-site researchers observed consultations, documented them carefully and analysed them for recurring practices in the sense of behavioural patterns. Video or voice recording of the real interaction between the client and consultant was forbidden due to privacy concerns. The consultations lasted for about 50 minutes. During that time the observers focused on the use of material and technology as well as the physical environment of the interaction. Their field notes included chronological listing of the participants' actions, as well as an exact description of interaction supported the usage of available artefacts including simple drawings of physical actions. The exact notes enabled subsequent reconstruction of the behavioural patterns and allowed discussion of them with the consultants and clients in interviews.

Following the consultation, the on-site researchers conducted interviews with the participants. First, the researchers approached the clients and requested a spontaneous interview. The clients were not offered any formal compensation for participating in the interview but the observers offered them a cup of coffee and a choice of sweets as a sign of gratitude. The client interviews took on average about 10-20 minutes and followed a semi-structured paradigm addressing primarily the elements clients liked and disliked about the consultation and their perspective on the use of the materials and systems during the consultation. The conversation with the clients allowed the identification of the most remarkable elements of the consultation in terms of the process: the clients referred to the process steps involved in the consultation but tended not to share their opinion freely if not explicitly asked by the interviewer. If asked explicitly, they were mostly provided general and/or positive answers about AA and about their consultants ('Yes, everything was OK, everything satisfactory…', 'I think, he wanted to get to know me, find out what I've done so far, and find something for me', 'I was surprised that it was helpful'). In particular, the clients did not reflect on the specific behavioural patterns of the consultants even if explicitly asked about them – they were simply responding that it was OK. Given that those behavioural patterns are at focal point of the study, this was disappointing. The clients' reactions might be an effect of the context: the interviews took place in the AA (though in a separate room) and directly after the consultation, which could make the interviewees hesitant to express criticisms in front of a stranger, even though the interviewers presented themselves explicitly as independent of AA and assured the clients of their strict anonymity. Also, 10 out of 38 clients did not take part in the interview while providing general excuses (no time, insufficient language skills, other appointments). While we believe them, such behaviour might also lead to self-selection bias in this part of our data. Because of the various reasons (generic focus, no reference to singular episodes, risk of self-selection bias), we have limit ourselves to including some general comments from the clients but have refrained from observing the interviews with them in the analysis of singular behavioural patterns which emerged during the use of IT.

Second, the researchers approached the consultants after visiting at least two of their consultations. The consultants were also offered a choice of sweets and beverages for their participation in the interview. The consultant interviews took about 45-50 minutes on average and made reflection on the observed behavioural patterns possible, an extended discussion on the characteristics of the systems and support material provided to them, as well as their goals and wishes. The interviews provided material to understand reasoning behind the observed behavioural patterns, as well as to characterize the consultants' own perception of their profession. Interviews with the clients and with the consultants were audio-recorded and transcribed according to the intelligent verbatim style (Cooper, 2014; Hadley, 2017). During transcription any information which could help identify the individuals was omitted and the original recordings were removed. Overall, the data consisted of field notes regarding the observed consultations (including notes on preparation and post-processing of the consultation) and transcriptions of the interviews. Table 1 summarizes the data collected prior to the analysis.

| Data collection type | Overall number | Involved consultants | Involved clients |
|---|---|---|---|
| Consultation observations | 38 | 14 | 38 |
| Interviews | 41 | 13 | 28 |

Table 1. Overview of the collected data prior to the data analysis.

During the data analysis, the data collected was coded using qualitative content analysis (Lamnek and Krell, 2016). In particular, the on-site researchers, supervised by three experienced researchers, coded the material along the lines of the identified behavioural patterns: they grouped the patterns, identified similarities and differences between and within the groups and finally selected material which explains the reasons for specific behaviour. The results present the identified behavioural patterns and exemplary quotes explaining the reasons behind them. The research team met regularly to control the progress and consistency of the data analysis and to discuss the identified patterns concerning the observed behaviour and the explanation for those patterns as expressed by the advisors in the interviews. The interviews with the clients turned out to be diffuse and illustrated a general attitude towards the employment consultation service e.g. they were not useful for the analysis of particular episodes or incidents which occurred during the service.

To make sure that the observed patterns and the identified reasoning were understood correctly and that the analysis offered reflects the actual thoughts and values of the consultants, we conducted a second round of interviews with selected consultants and organized a range of workshops involving consultants and their supervisors (who are or were employment consultants as well). We framed the workshops as moderated team discussions of about 2.5 hours long. The teams moderated by one on-site and one experienced researcher were approached the results by discussing the opinions and reasons concerning particular behavioural patterns. The workshops and post-analysis interviews were also used to collect opinions on the implication of the findings for the design of technology for employment consultations. The interviews and the workshops were audio-recorded and transcribed for a second round of analysis following the on lines described above. Table 2 summarizes the data collected in this phase of research.

| Data collection type | Overall number | Involved consultants | Involved supervisors |
|---|---|---|---|
| Interviews | 7 | 7 | 0 |
| Workshops | 4 | 17 | 3 |

Table 2. Overview of the data collected after the initial data analysis. The interviews and workshops were conducted to assure the accuracy and validity of the results.

The data collection and analysis process were focused on the observable behavioural patterns of the consultants during a consultation. This focus is consistent with the perspective on situation-specific practice as a nexus of overlapping and intersecting structures (organizational hierarchies, individual values, public and political discussions, structures provided by technology and material context, etc.). The analysis of the involved behavioural patterns enables the identification of the structures that are involved in shaping those patterns. During the interviews as well as subsequent workshops, it turned out that patterns which contradict or bypass intentions of the AA (as coded in written rule sets, popularized in training or implemented in their case and knowledge management system) are of particular interest to the employees of AA and are well-suited to making reasoning of the consultants more explicit. In fact, focusing on those specific episodes invoked a deep reflection on the consultants' side and provided an analytical advantage with regard to the discussed behavioural patterns but, at the same time, might be inappropriate for a comprehensive, holistic picture of the AA or of the consultants' job in its whole complexity. We refer the reader to other studies for this type of analysis (Bratteteig and Verne, 2012; Boulus-Rødje, 2018). Nevertheless, we will summarize the information on the complexity of the organizational and technological context of consultants' work which we obtained through the analysis of the interviews and the shadowing of the consultants (cf. section 4.1) to provide background for the analysis of the particular patterns observed in consultations (cf. section 4.2 and 4.3).

# 4 Results

The specific findings of this study, i.e., the identified coping strategies and the motives explaining them, are best understood against the organizational context of AA and against the overall picture of a consultation. This follows the principle of zooming-in on practices: we first provide the bird's eye view perspective and later fully approach the particular situation-specific behaviour. While the methodology applied throughout the study was concentrating on the latter aspect, many comments collected from the consultants and clients, as well as from the workshops with the supervisors, pointed to the organizational or even political aspects of employment consultancy. Also, the shadowing of the consultants provided material that contributes to the general understanding of the organizational context. Having understood the context, the coping behavioural patterns can be embraced in their full complexity.

## 4.1 Organizational Context

The employment consultation is one of the primary tasks at the AA. It is also one of the most visible tasks. The wider public connects AA with the obligation to attend regular meetings with a consultant during the unemployment period. Also, those consultations form the personal (face-to-face) interface of AA to its clients. For them, the consultant stands for the culture and the attitude of the AA as a whole. Many clients confirm  that AA has a negative connotation and word-of-mouth, but the particular experience was surprisingly positive and changed their perspective: '[It's important] that one meets on a level playing field, that you are being listened to, and that they are attending to your needs. (...) [Interviewer: Did you have other expectations?] Well, I know it from hearsay that you're being treated here as a second-class person. I've heard a lot but I was pleasantly surprised. (...) Now, I think, I will be adequately supported; it's what I expect now.' An advisor confirms that the consultation has potential to change what clients think or expect from the AA: 'It's about removing the anonymity from the perspective on the agency by becoming the face of the agency. I tell my clients: 'I am responsible for you. If you have any questions, feel free to contact me via Email or via the hotline''. Even though many clients perceive the AA through the lens of the consultation, AA is more than that. Indeed, consultations form an important part of the agency's support for the unemployed but other divisions in each of the AA's branches also contribute to the (re-)integration of unemployed individuals. The key players which are in contact with the employment consultant are listed in fig. 2.

Figure 2 also lists the core tools used within the AA, among others the core case and knowledge management system of the AA, VerBIS. VerBIS (Vermittlungs-, Beratungs- und Informationssystem; English: Placement, Counselling and Information System; (AA, 2013a)). VerBIS contains all information about all clients including job seeking people, unemployed people receiving monetary support, students who need consultation about possible job profiles, etc. No matter where in Germany they contact the AA, the employees of the AA can find information about them in VerBIS. At the same time, VerBIS is intended to provide a standardized way of collecting personal information, e.g. consultants (and other employees of AA) receive a structured characterization of each person's profile. VerBIS also imposes a range of regulatory constraints as it was developed to reflect the law in general and the duties of AA that follow from its role in the unemployment politics, as well as the AA's internal rules. From the organizational perspective, VerBIS forms the core technological resource of the AA and is used by most AA divisions. Other systems that are in use at the AA will be described in more detail along with their usage scenario from the perspective of a consultant.

Each consultant is a member of a consultation team. This team consists of all job consultants as well as the supervisor. Other consultants are important contacts for daily issues e.g. coordination and allocation of cases, intermediate help with technical, legal, or process-related issues, or temporary replacement in case of disease. Depending on the task at hand, they use a mix of communication tools involving face-to-face contact or phone calls (at meetings or for getting quick answers to easy questions), the general schedule management for coordination tasks, VerBIS to share information about specific cases (e.g., during sick leave), and email for a wide variety of tasks. The consultants' supervisor supports their team with similar issues, but additionally takes on quality management tasks such as work shadowing (visits to consultations of a given consultant), providing feedback or facilitating feedback within the team, as well as co-ordinating the training activities for their team. Contacts between the supervisor and their team rely on the same mix of communication tools. The supervisor also forms the formal point of contact to the managers of other divisions within the AA, higher levels of the hierarchy (the manager of a branch), and to the central organs of the AA.

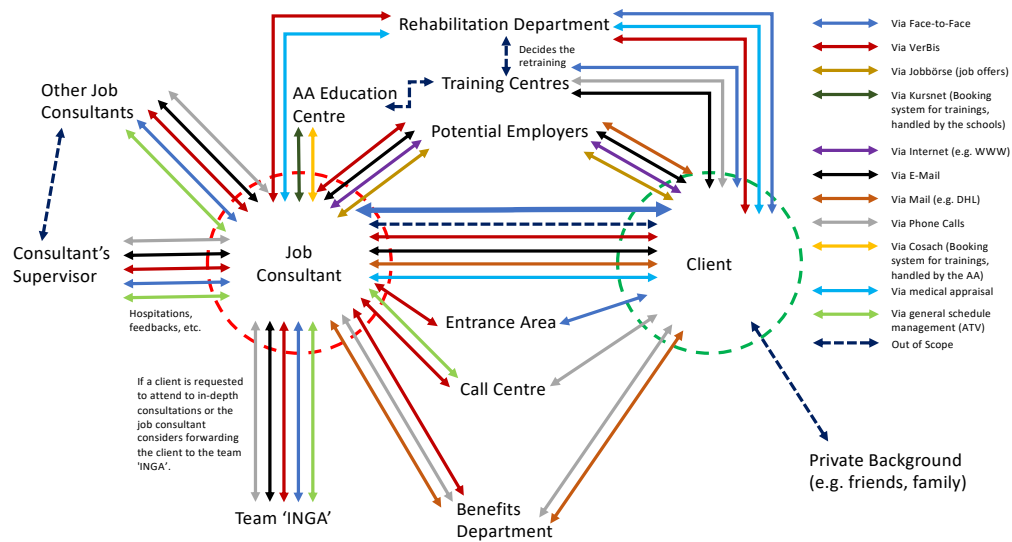

Figure 2. The organizational context of a consultant at AA, including the relevant stakeholders and the communication/collaboration tools used for contacts between them.

During their work on a particular unemployment case, the consultants need to collaborate with other divisions depending on the specific situation of the unemployed individual. For instance, if a client cannot work in their original profession because of any medical issue, the consultant is required to contact the Rehabilitation Department. Practically, if the consultant ticks the appropriate box in the VerBIS and provides the most necessary information (e.g., type of limitations experienced by the client), they will launch a process at the Rehabilitation Department. The Rehabilitation Department will co-ordinate the medical appraisal process for the client and approve a re-training, which will then be co-ordinated by the employment consultant. The contacts between the employment consultant and the Rehabilitation Department are highly structured and rely on VerBIS., The copy of the medical appraisal reaches the employment consultant in various ways (e.g., internal post or email).

In many cases, the clients require an additional training. It may be because of their health condition, when they need to learn a new profession. However, more often they require improvement of skills relevant for their profession or for the labour market in general – they might benefit from being brought up-to-date in self-presentation or self-organization skills as well as an update of skills concerning a new software package typically used in their profession. Many of the courses can be booked based on the agreement between the consultant and the client, without involvement of external units. Only the expensive trainings (e.g., learning a new profession) are subject to additional agreement from, e.g., the Rehabilitation Department. The list of the available courses is maintained by the Education Centre which collaborates with various Training Centres cooperating with the AA (e.g., regional centres for vocational training). They offer access to those courses via a separate system, Cosach. However, the schools and training centres can also list courses on their own in another system provided by the Education Centre, Kursnet. Both systems, Kursnet and Cosach can be used by the consultants to book training offers for their clients. Even though arranging a training opportunity to improve labour market chances for their clients is an important part of the case management, the process has not been implemented as a part of VerBIS.

If the client is subject to an unemployment allowance, registration with the AA will launch a process at the Benefits Department. This division of AA approves material and monetary allowances for the clients. While the exact internal processes of the Benefits Department are out of the scope of the current study, much of the information collected by the consultants in their consultations or otherwise provided by the client are related to the unemployment allowance processes. VerBIS is used to collect information about the client's attendance at the consultations, about their job seeking efforts, about the agreements made between them and the consultant, about their attendance to the compulsory training courses, about their health issues (e.g., doctor's confirmation of a sick leave when unemployed), and about their curriculum vitae including recent salaries. The Benefits Department has access to this information when deciding about allowances and some information in VerBIS launches specific processes in this Department (e.g., repeated unconfirmed absence at a consultation appointment). Here, VerBIS, as the main case and knowledge management system of the AA, plays its role as a documentation and control instrument. Phone calls to the consultant and occasionally to the client are used to clarify specific issues and to supplement the information collected in Ver-BIS. Internal mail is used to exchange supplementary documents between the consultant and the Benefits Department. Mail is used to inform the client about the decisions considering monetary and material allowances.

The ultimate goal of the re-integration programme of AA, of which the consultation is a part, is to lead the client back to employment. Potential employers play a key role here. VerBIS includes information about the employers in the region and about their domains and personnel demands. Additionally, the Jobbörse includes specific job offerings posted by the potential employers. Clients can access the Jobbörse on their own but this tool also gets used during a consultation, in parallel to the information collected in VerBIS, to support the ad-hoc job search activities. Because neither of the tools provides all the information about the potential employer, consultants often rely on external Internet tools, such as Google Maps, to locate separate branches of a company in the region. Consultants sporadically contact potential employers via email or phone to confirm information listed in a job offer, to facilitate the first contact between the client and the company, or to check whether a client contacted this company as agreed. Consultants do not have direct contact with the potential employers on a regular basis but they derive their knowledge about the them from online sources and from the information provided in VerBIS or in Jobbörse, this is seen as a challenge to their efforts in finding an optimal match between the client and a potential employer.

Apart from regular consultations and additional ways of contact between the client and their assigned consultant (involving VerBIS, Email, Mail, etc.), AA offers additional service channels for the unemployed: a Call Centre and the Entrance Area. They have supporting character and do not replace the consultation. Clients tend to contact the Call Centre if they are unsure about their appointment (e.g., when and where it takes place or if it is obligatory), about the status of their allowance decision (e.g., when they can expect the money to arrive in their bank account), or about the status of their agreement with the consultant (e.g., how many more applications they need to write and have confirmed). The Call Centre uses information collected in VerBIS and in the schedule management tool to answer those questions. In the Entrance Area of an AA branch, the client is prepared for the consultation (especially for the first). Even though clients are requested to enter their data in the VerBIS prior to the first visit to the AA, many of them do not do so – one of them explains: 'I tried that but because the Internet connection at home is not so good [laughs] (…) I honestly think that the application on the Internet is complicated to fill in'. In such cases, the frontline employees in the Entrance Area start entering the information together with the client, otherwise they check the information already entered. They also sporadically engage in triage activities, when clients without appointments arrive at the AA to make an appointment or even to have an ad-hoc consultation. Nevertheless, the primary role of the Entrance Area is to prepare the client with the necessary information about them for the planned consultation. However, because the time they have available per client varies greatly, the results of those preparations might be beneath the necessary standards – some clients arrived just before the planned consultation and did not prepare the necessary information while others are well prepared and arrive early. If the preparations could not be completed before the consultation, the time during consultation might be allocated for the collection of the key data and instruction to the client to complement and then complete this data after the consultation.

After the initial consultation, some clients with highly specific needs or in a demanding situation, who nevertheless want to find a job and integrate in the labour market, are classified as candidates for the internal holistic integration service (germ. *interne ganzheitliche Integrationsberatung*, INGA). These clients have no qualifications, no language skills, are homeless and without a formal address, are addicted to drugs, have experienced harassment or mobbing, were involved in crime, etc. etc. This service is offered at some dedicated branches of the AA and involves highly specialized consultants prepared to offer a holistic service. INGA combines labour market integration measures with these social welfare measures which are not covered by other national or local institutions. Some clients, e.g., homeless people, do not even attend an initial consultation with an employment consultant but are directed immediately to INGA. In other cases, the necessity to direct clients to INGA emerges during the consultation process, e.g., if the client refuses to apply for specific positions because of mobbing experiences. In such cases, the employment consultant will seek face-to-face contact with the members of the INGA team to discuss and transfer the case. The information about the individual will be provided in VerBIS. Additional questions are answered via email or phone.

The analysis of the organizational embedment of the employment consultants explicates the high complexity of their work. In particular, it shows that they need to approach various internal or external stakeholders depending on the characteristics of a specific case. This involves making decisions having an essential impact on the life of the client: is it necessary for them to change their profession or is it enough to improve their current training, do they require an holistic plan to participate in the labour market or are informal pressure and the available measures enough to motivate them, is it better for the client to attend a vocational course before applying for a specific job or should they just take a chance? Those and many more questions describe the consultants' area of impact. Even though they are deliberately separated from material and monetary decisions and have access to colleagues from the INGA team or the Rehabilitation Department for further exploration of a case, they are involved in complex decisions. Even more, they are responsible for framing the decisions as formal agreements with the client and then to check how the clients proceed with the agreement. The tools offered to the consultant are primarily designed for sharing information, orchestrating processes within AA, and for checking. Even though VerBIS covers the majority of these processes and information flows, other systems explicitly made to create a parallel structure (e.g., Jobbörse) or to replace VerBIS in specific domains (e.g., Cosach). Furthermore, WWW, email, phone calls, mail, internal mail, and other office software extends the information ecology in terms of tools and technology. However, there are no tools with the specific intention of supporting the decisions of the consultant or their face-to-face interaction with the client. This puts a large pressure on the consultation itself, which is the time when most decisions are made and agreed upon.

During the consultation, the consultant collects information about the client, systematizes and confirms information already collected in the system, as well as agreeing with the client on subsequent steps. During the consultation, the consultants are subject to a range of limitations and obligations. First, VerBIS imposes various technical and regulatory constraints. The technical limitations follow from the envisioned input and output modes, as well as from the envisioned functionality. For instance, VerBIS, was designed as single-user desktop software to be run by the employees of AA. VerBIS offers a separate view for the client as a web application, but this one is only available to clients and requires a client's personal password. Because this view includes information on the monetary support which the client receives from the AA, the consultant, who should not know about the allowance process, should not use this view during consultation. There is no dedicated version of VerBIS to be used during consultations. Second, there are regulatory constraints implemented via training and in coded in consultation guidelines. Many of those rules prescribe how they should use VerBIS. For instance, there are clear indications about what the consultants should document and in what way or how they should treat the data in the system. Of course, there are other regulations as well (e.g., what should and what should not be negotiated with the clients). We considered the rules in our analysis as far as they affected the use of IT. Above all, we can identify (1) technical limitations that result from the functional scope or the architecture of the VerBIS, as well as (2) regulatory obligations, which are in part directly implemented in the system and in part prescribed in the policy of AA on how to employ the system.

The on-site researchers attended the so-called initial consultations. The initial interview, as prescribed by AA, should proceed as follows: When citizens register as unemployed, they are invited to the initial consultation. Before this consultation they are sent various brochures by email about unemployment and with a request to fill in a questionnaire with details of their life, educational background, etc. The answers are stored in VerBIS and should support the consultants in preparing the initial interview. However, as mentioned above, some clients do not complete this task and require help from the Entrance Area employees. During the initial interview the job consultants are encouraged to build up trust and a solid relationship with the client. In parallel, they should carry out a profiling together with the cli-ents. This involves collecting data on qualifications, skills and abilities and assessing and documenting characteristics such as punctuality, accuracy, interest, etc. In addition, the client's personal situation is examined in order to identify possible obstacles or advantages regarding the job placement. When all the data is available, the job consultant decides what to do: they might identify a set of missing skills and agree with the client on a training programme or they might identify potential health problems and contact the Rehabilitation Department as described above. Given that the client is healthy and no essential skills are missing, consultant and client seek suitable jobs or appropriate employers to be contacted; when doing so, they compare the profile data with the job requirements. The clients should apply for the appropriate positions after the consultation. Finally, the client and the consultant prepare an integration agreement including information on when and what the client has to do. These integration agreements are required by law and serve as a basis for evaluating the customers' commitment to take up a new job.

The above description of an initial consultation follows the process of conducting a consultation. However, the burden of making complex and making an impact on decisions in a collaborative manner is barely touched upon. Many decisions to be made during a consultation have far reaching, formal consequences and should result in an agreement. In this situation, the consultants establish practices that, to a certain extent, alter or complement the formal vision of the initial consultation. In the following we attend to those behavioural patterns, which lie at the heart of this study.

## 4.2 Coping with technical limitations

***Turning the monitor to and for the client***: During a consultation, the consultant sometimes turns the monitor towards the client. As presented in fig. 1, a flat-screen monitor is normally positioned at the top of the table and can be rotated back and forth. In its default position, the monitor faces only the consultant and the client cannot see the contents displayed. Consultants rotate the monitor to and for the client once or twice during the consultation. We observed instances of this behavioural pattern when the parties discussed and extended the client's curriculum vitae, analysed his/her strengths, or when they both engaged in searching adequate job offerings. The physical action of turning monitor was accompanied by explicit or implicit requests from the consultant to focus on the display such as "let's have a look" or "you see, there are…" and deictic utterances. For instance, when discussing the curriculum vitae while having the monitor turned to the client, the consultants often preferred spatial prepositions ("down here", "below", "above") but while discussing the same topic without sharing the display, they use content-related markers ("before", "after"). The consultants tended to use the mouse cursor as a pointing device but also used a physical reference, e.g., a pen or their finger as a pointer, when they engaged in longer explanations. Whenever the activity was about collecting information, they hardly ever used physical pointers and stayed only with the mouse. Turning the screen towards the client marked a shift in activity and the content of the screen became a spatial resource for the consultation. As presented in Table 3, consultants explain their behavioural pattern while referring to various dimensions of a consultation including transparency, relationship to or understanding of the client.

| Reasons for turning the monitor towards the client | Statements of the consultants |
|---|---|
| a. to improve transparency | "when I need to collect further data (…) Otherwise, I key in the data ten minutes long and the client doesn't understand why. " |
| b. to improve relation to the client | "I claim that I am closer to the client that way than if I am [just] looking at the screen" |
| c. to improve understanding | "if I have to explain difficult things (…) then I turn the monitor so that they understand." |

Table 3. Selection of frequent reasons for turning the monitor towards the client.

However, the consultants turned the monitor back to the default position shortly after completing the activity for which they had wanted to share the screen. We observed that the instances of these behavioural patterns correlated with such actions as: switching to another application or to a new screen of the active application, switching to a free-text input field or opening a new document. Interestingly, turning the monitor back to the consultant was not accompanied by any specific verbal comment or excuse. This suggests that the monitor when turned towards the consultant is considered in a default situation and neither explicit nor implicit explanations are necessary. In multiple cases we observed that while the consultant was turning the monitor back, the clients changed their bearing: they sat up straight or leaned against the chair and gazed at the advisor. Turning the monitor away from the client marks a shift activity, which is reflected in the reaction of the client. The consultants provide a range of reasons why they turn the screen back, so that the client cannot see the content displayed. Some are driven by rules (to protect sensitive data), while others consider their own comfort or the comfort of the client. Table 4 summarizes the reasons which have been identified.

| Reasons for turning the monitor away from the client | Statements of the consultants |
|---|---|
| a. to protect sensitive data | 'The client is not allowed to see the internal notes' |
| b. to avoid questions | 'Because there are a lot of internal company matters, also for the social benefits division, which then grants costs. It would bring them nothing and would probably mean more work for us, because clients would ask questions about things they cannot and need not understand' |
| c. to increase our own physical comfort | 'I do not know how the client finds it – it is, of course, sometimes uncomfortable for both sides to look at the monitor from around the corner, especially when I have to write something' |
| d. to avoid embarrassment | '[according to regulations] the job openings should not include any information; a job offer should be free of statements that the company does not want women, does not want Turks and wants only people younger than 25, it shouldn't all be there; but it happens and then I go over it very quickly because I am so embarrassed' |
| e. to avoid confusion | 'The biggest problem is the different interface, the clients work with „Jobbörse", we, the consultants, work with "VerBIS." (…) The screen is different and only the data are identical (…) I think I can't turn the screen and tell the client: «Look, here you can find information» and back home he gets a different screen with a different look and feel.' |
| f. to avoid disrespect towards the client | 'One should actually do profiling in every conversation, together with the customer; but I have never done this, because when I read out, for example, 'there's need for action concerning <u>intellectual capacity</u> or self-initiative…'; you know what I mean; I do not do this with the customer, I just do it by myself' |

Table 4. Selection of frequent reasons for turning the monitor away from the client.

***Using the Internet-wide search and online resources***: Even though VerBIS offers access to job offerings and training opportunities for the clients, the consultants often switch to non-specific tools such as Google Search, Google Maps, or a particular website in which they know it includes the necessary information. They do so even though the official documents of AA do not mention the use of external knowledge sources. The consultants simply open a standard web browser on their desktop and put other applications in the background. They mostly communicate this fact to the client when initiating the action "let's check with Google" or "there is a database of X on the towns homepage". This action often involves turning the screen towards the client and inviting him/her to join in. When reflecting upon the use of external resources, the consultants refer to the shortcomings of VerBIS or content collected therein such as missing contact data, fragmented and distributed information, or insufficient functionality compared to standard consumer tools. However, their reasoning for why it is reasonable to use external, online sources of information during consultations goes beyond that. The consultants see it as a way of providing better or more creative solutions for the client and to empower him/her. Table 5 lists the reasons behind integrating external tools in consultations.

| Reasons for using Internet-wide search and external resources | Statements of the consultants |
|---|---|
| a. to link pieces of data | 'If I notice: 'aha, the client so and so lives in Gröbingen', and if I see an advertisement that fits him and (…) if I knew approximately how far this place is from there [potential workplace] or similar small pieces of information, I would not have to search and think so much' |
| b. to improve understanding on the client's side | 'If the client complains that she can't apply for the job because it is too far away or so, (…) I use a route planner to see and illustrate how far it is from her home.' |
| c. to accommodate free exploration and search for solutions | 'I need Google, yes, if I think in some direction, I contrive ways and means, and I tell the customer it's about improvising too; then we can be a little creative together; I try to look out of the box and then also have ideas: as to where to start, where you can find out more, where you get additional data' |
| d. to complement incomplete information | 'What I did not like about this particular consultation was that I spent so much time searching for this German course; and this was just because I couldn't open that link in that email!' |
| e. to empower clients | 'I show the client how it is possible he can help himself, an aid for self-aid; and not that I hand him everything on a plate; so that I can also say to him: 'do some research on the internet for XYZ'' |

Table 5. Selection of frequent reasons for using Internet-wide search and other external resources.

***Combining paper and computer:*** The consultants are well aware of the limitations of the system provided for them. This becomes more obvious when one considers the efforts they make to complement the ability of VerBIS with other systems (as above) or with paper. Four behavioural patterns dominate how consultants introduce other materials than computer in their consultations. Firstly, they use standard brochures provided by the AA or local agencies – they possess access to all standard brochures in their offices and know well what information can be found where. Therefore, when using a brochure, they almost intuitively reach out for the right document and open it on the necessary page. Upon presenting the information in the brochure, they often give it to the client and make a specific note in the documentation. Secondly, consultants use a selection of materials which they prepared for themselves (e.g., print outs from online sources) – they use them in a similar manner as the official brochures. Thirdly, consultants prepare print outs from the documentation they created together with the client during the consultation or the job offerings they selected. Some consultants even print two copies of it and let the client sign one of them, which they then put into their archives. Consultants prepare the printouts towards the end of the consultation and do not discuss its contents with the client again but sometimes they add markings which help to identifying key aspects. Fourth, consultants sometimes engage in drawing or writing down important information on paper. For this, they put a piece of paper in the middle of the table, between themselves and the client, and start drawing or adding information to a printout. This can occur at any moment during the consultation. So, the consultants include paper-based materials in almost every consultation. Table 6 provides an overview of the most frequent reasons for those behavioural patterns. Apparently, paper-based materials outperform the online tools in terms of supporting understanding, focus, and legality.

| Reasons for combining paper and computer | Statements of the consultants |
|---|---|
| a. to improve understanding | 'I like to take a sheet of paper and do the pros and cons if there are a few aspects positive and some negative (…) What I always include with this is that [getting the statistics on integration opportunities after x months unemployment; also used by other consultants]. The statistics are helping people to understand things. I also have something laminated, I personally use a lot of the laminated material, for example, for the business-political goals [of the AA].' |
| b. to reduce information overload | 'So, I notice that if you send the customer 27 different papers and text modules, they do not read them, but if I write down or print only 3 things then they read it.' |
| c. to guarantee legality | 'Email has no legal consequence, so all job proposals, on which a client must apply, must be sent by traditional post.' |

Table 6. Selection of frequent reasons for augmenting the consultation with paper-based materials.

### 4.3   Coping with regulatory obligations

Consultants are required by law and by internal AA's rules to document their contacts with the clients. The required documentation might be overwhelming especially in the first consultation – the consultants need to document data collected about the client, his background, his education, and his personal context. The extended documentation should support the binding character of agreements between the parties and be a guarantee for standardization and compatibility of processes, so that a client can switch to another consultant if necessary. A client might also request access to all of his data, make claims against how the agency treats him or simply make allegations which need to be compared with the documentation. Of course, this creates pressure on the consultants as one of them illustrates: 'When a new client starts, you get a document and you just pass it on to the social benefits department. You document this. The client contacts you without appointment, you document it. They call or do anything else, you document it; everything has to be documented; it is not only the once during the first conversation but also in the follow-up talks. (...) We write a diary, so to speak. It's like a diary for 300 people'. Even though the consultants engage primarily in consultations, as a caseworker managing hundreds of unemployed individuals, they need to track interaction with each of them, even though the documentation is hardly ever used by consultants themselves or their colleagues but remains a formal and regulatory duty.

VerBIS was designed in accordance with the regulatory expectations and provides space for entering structured and unstructured data. The user interface consists of links, buttons, and text fields of various size, of which some are required while others can be left empty. Figure 3 shows the interface used to create a single work-experience entry in the curriculum vitae of the client. Given that a client might have extended work experience with multiple companies and at various positions, the consultant might need to fill out this form several times during a consultation. Importantly, work experience is just a single part of the curriculum vitae and updating/extending the curriculum vitae should be just a part of the consultation.

In fact, it can easily become more data that needs to be inserted into the computer. Since the other systems provided for the consultants are not linked properly with VerBIS and automated transfer of data does not work, the consultant sometimes needs to enter the same data into two systems (for instance, VerBIS and a separate case management system for re-integration services to book vocational training). So, the consultants face regulatory requirements to provide extensive documentation but are provided with tools that do not support them in swiftly creating this documentation. On the contrary, the systems require manual input of textual information and so make the documentation even more inefficient.

Figure 3. An exemplary screen from the VerBIS (AA, 2018b) used to enter details about a single entry in the curriculum vitae including three sections: general information, details about the type of the entry, and details about his/her occupation.

In this situation, the employment consultants developed ways to cope with the usability issues of the system on the one side and the regulatory duties on the other. We observed two ways of how consultants deal with the regulatory expectations. To reduce the effort of documentation during the consultation, they either advance or postpone its the creation. To reduce the overall effort of documentation (within and outside the consultation), they select what information should be documented and what not, despite their regulatory duty. Additionally, even if AA puts much effort into creating a complete documentation, there is only some limited information available about some clients in the system. In other cases, consultants might lack time to leverage the value of the whole documentation.

***Advancing or postponing the documentation***: We observed three behavioural patterns, which help the consultants reduce the effort of documentation during a consultation. Firstly, consultants created their own individual collections of frequently used paragraphs, which they copy from local files, where they store the text snippets in VerBIS and other systems. Secondly, consultants use only keywords to make notes (either on a piece of paper or directly into the system) and they complete the documentation afterwards. Thirdly, consultants fill out only parts of the documentation, mostly examples of repetitive sections, together with the client and ask the client to complete the documentation afterwards. So, the consultants cope with the time issues and ineffective data input by moving the typing outside the consultation: either by advancing the note taking activity by preparing text snippets they might use in the future or by postponing documentation to the time after the consultation. Table 7 gives an impression of the reasons behind consultants' ways for reducing the time for typing and documenting during the consultation.

| Reasons for advancing and postponing the documentation | Statements of the consultants |
|---|---|
| a. to save time for other tasks | 'Usually we do the profiling, the job seeking with the client and the notes about our conversation: what we explained and assessed, the rules we talked about. And we have to take those notes but all this writing we do afterwards.' |
| b. to improve quality of the documentation | 'When I talk in the consultation, I write down keywords (…) I write down the specific keywords and can make my documentation afterwards, with all the information I find important (…) it would sometimes be more useful, if my keywords were a bit more detailed' |
| c. to improve quality of the answers from the client | 'We, consultants, give them [the clients] the list and tell them to do it at home because we think that the customers should think carefully when answering, and this is not possible in the first consultation.' |
| d. to enhance efficiency | 'Because it simply has to do with time management so that I do not need to write the sentence daily if I can write it once and then copy it over.' |

Table 7. Selection of frequent reasons for advancing and postponing the documentation.

***Documenting only 'important things'***: According to the internal regulations, the consultants should document each interaction with a client. However, apart from interacting with the consultant during the consultations, clients make use of emails and the telephone to contact their consultant, e.g., when they interact with the system to complete their curriculum vitae and need some assistance. AA's official regulation would require the consultant to document such contact, however some advisors feel free to decide when documentation makes sense or not. Firstly, they choose which interaction to document and which not, based on whether the question is related to the unemployment or it just affects the co-ordination. Secondly, they decide about the effort to be put into the documentation depending on whether they expect that the client to stay unemployed or not. This affects in particular the integration agreement, which is a form of documentation – it records that the consultant and the client agreed on an integration strategy: 'I often do not do it in the first consultation; if there's a so-called job-to-job client, where it's already clear where the story goes (…) then I do not do it'. In reality, VerBIS affords extensive documentation about a client, but the consultants try to find an appropriate level of detail and amount of documentation based on the predicted use and future value of their notes and the produced documents. Table 8 gives an impression of the reasons behind selective documentation:

| Reasons for selective documentation | Statements of the consultants |
|---|---|
| a. to fulfil the perceived formal duties | 'For example, there are colleagues who do not make a note in VerBIS when the customer just visited them without appointment. Others document this word-by-word. (…) Some do it and some do not, some give it in detail, some do not' |
| b. to accommodate the client's right to see the documentation | 'I definitely document important things. (…) Because it is also the case that the client has the right to see everything that is stored here about him/her, so one cannot of course note down any private opinions or anything negative; but of course there are wordings or phrases from which at least I can recall things in the next conversation or perhaps even for a colleague, who later joins the case and can also draw conclusions from it' |
| d. to support subsequent consultations | 'If I maintained the record well, then I saw everything at a glance [at the beginning of a consultation]. Who's that, what's his/ job, where does he/she live, does he/she have a herniated disc, does he/she have a car.' |
| e. to accommodate unclear guidelines | 'It's like this: What can I document, what am I allowed to do? It is visible to others, for the customer, too. What about privacy? I cannot write now: 'customer got angry or is aggressive'. These are things where I reach my limits and think: Can I do that? Must I? Can I?' |

Table 8. Selection of frequent reasons for selective documentation.

***Getting a general picture from limited data:*** Given the effort involved in documenting the interactions with clients, one expects the system to provide an extended data set for each individual. But this is not always the case; normally the documentation of the clients first-visit is especially scarce. There is a regulation that each client has to fill out a form concerning his CV. Even though most clients are requested to fill out information online before attending the employment consultation, only a few keep to that. Clients often feel overwhelmed by the amount of information and number of requests they receive before the first visit to the employment agency, and, while ignoring potential sanctions, they appear at the first consultation without having used the self-service to input the data into VerBIS. Instead of sanctioning them, consultants continue with whatever they have and prepare for the consultation in a limited manner. And, they do not see that as a disadvantage: 'I've found out that, quite often, the preparation did not actually make sense because the person sitting in front of you described something quite different than that which was in the CV or in the previous remarks'. The previous documentation and the data inserted by the client play a secondary role for a consultation.

Independent of whether a client profile embraces a lot of data or is almost empty, consultants claim that focusing on few central data points has its advantages. Some argue that a good CV provides the right information in a very short time, thus providing the right time-to-inform ratio: 'So it's enough for me if the CV is well filled then I can already make a good picture of what is approaching me (…) It does not really take much time; so in 5 minutes you know just enough'. Other consultants do not consider even this short preparation necessary and argue, it is even better not to read the collected documentation to prevent prejudice: 'I often do not look at it before the first conversation, so I can just form my own opinion and then I only look at it afterwards'. The regulations prescribe documentation, among others, to support preparation for the consultation. It seems that the guidelines follow the principle of *the more, the better*. However, this requirement is overwhelming not only for the consultants, so that they document only the important things, but also for the clients, who do not consider the requests from AA. Nevertheless, consultants' established practices allow them to cope with too little or too much information by focusing on just enough data.

## 4.4 Consultants' general goals and drivers

Whereas the coping patterns each has its own singular rationale and explanation, the interviews provided us with general job objectives and triggers for consultants' behaviour. Consultants intend to support a sustainable integration in the job market: '[My objective is] the integration into work. But under the condition that it is a sustainable integration for him and is also an integration that is in his favour'. This might only be possible, when the client provides the relevant information about him/herself: 'well, I need to know things and if they do not tell it I have to ask them outright (...) if I do not know their needs, I cannot do anything for them'. Additionally, some consultants see information provision as a central aspect of their job: 'when it comes to unemployment benefits, there are a lot of rules involved, and I have to transfer them somehow [to the client]'. Nevertheless, they also see emotional support and the necessity to bolster up the client a bit as a way to actually help someone leave the state of unemployment 'I think it's always very important [to bolster up the clients] because often, they come and think they won't have a chance anymore or they are afraid of the AA. Ideally, they leave with a good feeling and are optimistic'. The general job objectives identified by the consultants are in line with the explanations they use when they argue for their coping strategies. In particular, they make clear why so many coping strategies are about freeing up some time between the numerous regulatory and system-related duties: the job objectives identified by the consultants go beyond processing many cases in formally perfect manner and are about providing client-centred support.

# 5 Discussion

## 5.1 The overestimated role of technology

The discourse on street-level bureaucracy brought about a range of sweeping and generic statements about the role of technology in frontline interaction. In particular, it saw technology as the game-changing force spreading across agencies to reduce the discretion of a single employee to a minimum and, eventually, replace them (Bovens and Zouridis, 2002; Lipsky, 2010). However, it also noticed that the pace at which changes have occurred stagnated leading to the proliferation of screen-level bureaucracy across agencies while attributing it to the deficiencies of the self-service technology or general societal or economic contexts (Hupe and Buffat, 2014; Hansen et al., 2018). Whereas the literature provided valuable and essential insights, for instance on the popularity of particular interaction channels (Scott, 2004; OECD, 2015; Hansen et al., 2018), it kept positioning technology as a driver by itself with some exceptions that consider the design of a specific system (Kirk, 2000; Giesbrecht et al., 2016a). Especially the study of employment agencies lacked studies that report on the usage of the case and knowledge management systems in the work context. While following the tradition of practice research (Orlikowski and Barley, 2001; Orlikowski, 2008; Schmidt, 2011; Schmidt, 2014; Schmidt, 2018) and ethno-methodologically-informed workplace studies (Heath and Luff, 2000; Luff et al., 2000), this manuscript opens up the impact of technology on the collaborative practices in employment consultations.

As an answer to the first research question *(How do employment consultants cope with inconsistencies between IT structures and other demands during an unemployment consultation?)*, this manuscript presents a whole set of various practices which consultants employ to balance up the impact of conflicting structures. It particularly points to the frontline employee as the entity in control of how much and what sort of technology diffuses at the interface between the citizen and the agency. Technology itself poses a curtailment up to a certain extent, the consultants let it curtail their actions and it poses an enablement to an extent and the consultants employ it to improve their service. The consultants actively overcome the curtailing elements by employing external tools, adjusting the amount of computer use during the consultation or by employing the client's own resources when encouraging them to search for information. Those combinations require creativity and experience beyond what the systems offer. As claimed in the curtailment argument (Bovens and Zouridis, 2002; Snellen, 2002; Lipsky, 2010; Buffat, 2015), the consultants gave up some authority – they cannot decide freely on sanctions against the client or the acceptance of certifications. The curtailment comes from the rules implemented into the software and, primarily, from the extensive documentation duties which require legal argumentation for each decision. However, consultants still have the freedom of effective use and efficiently management of information sources and resources; they can decide when and which resources to use, e.g., what to document or what to search for. Without the consultant's help, a client may lack the information about accessible or available resources or ways to use them (e.g., formulating an effective search query that describes their situation or their wishes) or they may be overwhelmed by it if they do not know what is relevant enough to document. As a result, it is the creativity and expertise following the handling of many cases that helps the consultants overcome the limitations the technology provided and regulations.

However, technology contributes to the enablement of consultants as well (Bovens and Zouridis, 2002; Jorna and Wagenaar, 2007; Buffat, 2015). With such a simple measure as turning the monitor, the consultants turn a single-user system into a collaboration tool with the capacity for collaborative searching (Greatbatch et al., 1993; Greatbatch et al., 1995; Kirk, 2000). With reviewing partial information provided by the system, they can establish a picture of the person they are talking to. This empowers the consultant to interact with the clients in a more personal way, along with their desired job objectives. At the same time, the results make clear that the technology can impact the frontline consultation only as far as the consultant allows for this influence. Thus, the street-level discourse overemphasized the influence of technology compared to the role of organizational policy (e.g., does the management use the controlling functions in the system to oversee the employees?) and, finally, to the actual power of the frontline employees, who may simply overcome the guidelines on system usage or employ alternative systems, which they feel are more appropriate or more empowering for themselves or the client. However, the results provide insights on how technology curtails and enables the frontline personnel and points out that the two tendencies can exist side by side.

To provide an answer to the second research question *(What are the reasons behind these specific coping behavioural strategies of employment consultants?)*, we add explicitly to the explanations provided by the consultants. Whereas previous studies on the use of desktop computers in consultations identified specific behavioural patterns typical for this setting (e.g., turning the screen back and forth or referring to external tools even though internal ones are provided) (Greatbatch et al., 1993; Greatbatch et al., 1995; Pearce et al., 2008; Pearce et al., 2012; Kilic et al., 2016), this article identifies the underlying motives of the consultants engaging in these practices. Those motives go beyond the design of the technology: They take into consideration the actual and the perceived or assumed needs of the clients, just as when the consultants explain that they turn the screen or use popular rather than AA-typical tools to help the client understand. They also take the client's assumed knowledge, competence or self-esteem into consideration, as to when the consultants want to shelter clients from confusing system terms or from depreciating content. They also envision client's future needs or questions, when they try to empower them by showing how to complete specific tasks with the tools available. In short, many of the behavioural patterns are seen by the consultants themselves as client-oriented and resonating with their general objectives to bolster up the client to such an extent that they can act assertively and collaboratively (rather than passively) with the AA. However, the consultants also refer to their personal beliefs about the quality of the consultations and how the rules and the technology curtail or enables them to deliver services of the quality they strive for.

While the consultants do not talk in categories of curtailment and enablement, they refer frequently to usefulness and efficiency. And so they document useful, *important* information and they retrieve useful, *meaningful* information, they reduce the amount of inefficient data input to make time for other activities and so on. This shows that the consultants feel rather confronted with too many ineffectual tasks that cost too much time. And this is their primary concern. Therefore, rather than discussing the role of technology as a principal driver of change, we call for a more action-oriented perspective, that sees technology as a tool to be embedded in existing and desired practices. In particular, we call for the design of systems that consider the complexity of frontline interaction with the unemployed, often vulnerable or insecure individuals. Even though there exist few notable examples which illustrate how to design and implement technology that empowers frontline employees (Scott, 2004; Letch and Carroll, 2007; Campbell et al., 2013; Giesbrecht et al., 2014; Giesbrecht et al., 2016b). We still lack a coherent, consistent body of knowledge about domain-independent guidance.

On the one hand, there are suggestions which have an intermediate character and are direct responses to the identified shortcomings in the AA. Firstly, VerBIS should be integrated with publicly available resources on the Internet – allowing the clients to re-use skills from their previous private and professional life rather than learning how to use a system just for the unemployment period; this would support the consultants at empowering the clients. Secondly, it should provide an easy documentation method – since it is enough when written documentation includes only the central facts, the other data could be voice-recorded on demand either during the consultation or afterwards. Thirdly, it should accommodate for a collaborative interaction space during the consultation – the consultant could so display singular windows or parts of the screen to the clients, without the necessity to hide sensitive information or turn the monitor back and forth. Fourthly, it should generate the integration agreement and other remarks automatically, based on the actions taken in the system – the consultant could offer a binding document as a product of the consultation even for clients who are expected to terminate their unemployment soon. The proposed improvements assume the current setting and the provided tools to remain within AA for the next few years.

On the other hand, the study provides a range of more general suggestions for the design of tools and systems used in public agencies, especially those handling complex social cases. Firstly, the findings imply that the consultants understand themselves to be social actors (as opposed to political actors or a supervisory body). In their explanations they refer to clients' needs or their own vision of a good consultant but leave out the political dimension of AA activity. This is the impression they want to pass over in their behavioural patterns. Research in bank advisory services has previously described a similar phenomenon and shows that systems whose design focuses on the impression (rather than on objective goals or tasks in a consultation) gets appropriated more easily and smoothly into a consultation and enables crucial social rituals to emerge (Dolata and Schwabe, 2017b; Dolata et al., 2019). Secondly, the case and knowledge management systems acknowledge the high relevance of asynchronous collaboration between the various stakeholders (cf. Figure 2) but ignore the fact that most of the complex collaborative decision making happens during the consultations between the consultants and the clients. It is not only the matter of a screen that cannot be shared easily but also the lack of effective tools - These tools make the decision process explicit and comprehensible for the client (improving the understanding of clients is the key reason for several of the observed behavioural patterns) or make possible a case-based comparison concerned with the effectiveness of specific decisions taken in the past. Previous research in various domains stresses the educational aspect of a consultation, e.g., in patient education (Greatbatch et al., 1995; Pearce et al., 2012), and other studies it is suggested how design can support those processes in other domains (Giesbrecht et al., 2014; Heinrich et al., 2014; Giesbrecht et al., 2015). Thirdly, the findings support the perception of data work as a possible collaborative activity between the client and the consultant (Fischer et al., 2017; Pine et al., 2018; Bossen et al., 2019). This leads to a comprehensive understanding of their situation. The fact that the various departments of AA use channels like face-to-face conversations or phone calls to contact the responsible consultant when treating a case shows that the decisions they are taking are more complex and require an holistic understanding of the client than it emerges from the documentation in VerBIS (independently of how exact, formal or informal it is). Thus, to support decision making within a consultation and beyond it, the development of the design of the tools should be focus on the comprehensive understanding of the cases rather than be standardized and highly structured but with fragmented documentation. The current study provides a range of findings showing that the transfer of suggestions from other consultation domains is necessary and needed. However, it also points to the difficulties and highly complex nature of case handling in employment agencies that require careful consideration of the various stakeholders and their identities in the socio-technical system.

## 5.2 The underestimated role of consultants

The personal job objectives of the consultants seem to be the key to understanding the role of technology in frontline interaction at public agencies. While previous research already suggested that there might be differences between the employment agencies and the consultants concerning the objectives of consultation services (Osiander and Steinke, 2011; Boockmann et al., 2013; Kupka and Osiander, 2017), this study provides a more detailed picture. The objectives of the consultants are nuanced and consider empowerment, empathy and sustainability to a larger extent than the regulatory framework; their statements point to the social competence as key driver of their behaviour. These job objectives establish an additional structure that forms the daily practice (Orlikowski, 2008), apart from the other structures and frameworks externally provided for them. The impact of the internal job understanding becomes explicit, when one compares the general objectives and the reasoning behind the particular practices. For instance, consultants omit formal language when they feel it could affect the client's self-esteem and thus contradict their objective to bolster up the client; they employ external tools and share their screen with the client to empower the client and to provide the most accurate and relevant information for them. To arrange for this, they also reduce the time necessary for formal tasks. In general, in their daily practice the consultants enact the structures available to them - be it through the organizational frame, the system or their own vision (Orlikowski and Barley, 2001; Orlikowski, 2008). We claim, they go even further – they mediate between the structures to maintain a balance between the expectations coming from various sources.

Consultants act as mediators between multiple structures when engaging in practices. They mediate between the client and the system, when they keep potentially depreciative or frustrating phrasing for themselves. They mediate between the system and their own needs when they focus on useful rather than comprehensive documentation. They mediate between various systems, when they manually transfer and adapt data or when they complement data from one system with information from another. Finally, they mediate between the macro-level context of the large organization, AA, and the micro-level context of a consultation when they move formal tasks to the post-processing or when they actively postpone the moment of creating the integration agreement. Therefore, in their role as mediator, the consultants fill the gaps between IT, regulatory frameworks and the situation or the client identified in past research (Bratteteig and Verne, 2012). Because the consultants work at an intersection of various expectations and formal frameworks, they are constantly involved in those coping strategies. Instead of making resources free for mediation, IT currently poses an additional source of pressure, which forces the consultant to fill more gaps. As a consequence of this, the consultants act not as a passive interface between the client and the system but as active and creative shapers of the client experience. In a sense, the screen-level bureaucrat turns into a gap-level consultant.

In this context, we want to stress the role of the seemingly insignificant material of behavioural patterns that consultants engage in when providing the consultation. In line with the multi-dimensional perspective on practices (Dourish, 2001; Orlikowski and Barley, 2001; Scollon, 2001; Orlikowski, 2008; Schmidt, 2014; Schmidt, 2018) and convinced that practices emerge at the cross section of various structures (material, organizational, political, social, etc.), we ask for attention to those performances. They carry and express the key values of the stakeholders involved and, especially in consultation situations, can strongly impact the perception of the whole organization: the consultant is frequently the only person a client will meet personally during their interaction with the agency. Even though the clients interviewed hardly ever referred to particular behavioural patterns, they often expressed the general feeling of being in good hands. Our call goes to the research community who, driven by the intention to understand the big picture and depict its complexity, overlooks the "small" behavioural patterns. But it is exactly those behavioural patterns which can act as a nexus of the complex whole and, thus, provide a relevant and more tangible entry point. On the other hand, this call also goes to the agencies and decision makers, who, driven by the political directives and organizational workflows, ignore the material aspect of the tools they provide for use by their employees. In particular the frontline employees, who represent the whole organization when interacting with clients, receive tools that barely fit the material and social conduct for their daily activities.

# 6 Limitations and conclusion

The results do not come without limitations. Firstly, the study is limited to two employment agencies out of 156 distributed across Germany. Extending the observation to other branches and agencies could support the validity of the results; extending the observation to other countries could provide an opportunity for a comparative study. Secondly, the opinions collected from the clients were mostly very general and rarely mirrored the singular behavioural patterns. A part of the problem was the limited time (they were asked for participation ad-hoc after the consultation and rarely had more than a couple of minutes); arranging an additional interview was impossible because the on-site researchers were not allowed to collect any data from the clients (by email, telephone) and because they were not allowed to materially compensate them for participation in the interview. According to legal advice from AA, clients would have to disclose such compensations to AA, and it would be deducted from their unemployment allowance. Collecting opinions from clients in a less rushed manner and with more compensation could make this source of knowledge more fruitful. Thirdly, the possibility to record consultations could contribute to an even more precise description of the coping strategies involving low-level features like particular phrases or non-verbal behavioural patterns employed by the consultants. Fourthly, the data collection and analysis methods were chosen in accordance with our primary scientific lens: material and situated practices in consultations. This data cannot be used to make implications about the overall structure of the AA but it is rich enough to depict the complexity and the inferences between consultations and other processes of AA as explained in Section 4.1. So, a range of limitations emerges because of the organizational, real-world context of the study and privacy concerns.

Even though many citizens connect a visit to a public agency with a fear of a negative outcome or even punishment, this vision seems to be overdone. As illustrated by the example of the public employment agency in Germany, AA, the consultant is likely to mediate the technical and legal limitations to offer empathetic support. Their internal motivation and job objectives are far away from the vision of a passive bureaucrat or a disciplinarian – they want to deal with clients' issues creatively and competently without an overhead on technical limitations or regulatory obligations. Consequently, they engage in various work-arounds, which become routine practices and, finally, form interrelated coping strategies. By illustrating these strategies, the current manuscript describes employment consultants as active and creative supporters for clients in a difficult situation. In light of the practices that have been identified, the consultants neither appear as arbitrary decision makers, thus contrasting with the vision of street-level bureaucracy (Lipsky, 2010), nor as a passive interface to a system and thus contrasting with the picture of a screen-level bureaucrat (Bovens and Zouridis, 2002). We argue, one cannot also see them either as a mixture between the two or as an in-between (Buffat, 2015; Hansen et al., 2018). Instead, they are constantly mediating between the forces that drive them in their daily practice. This perspective adds to the street-level bureaucracy discourse in eGovernment (Snellen, 2002; Landsbergen, 2004; Letch and Carroll, 2007; Lipsky, 2010): the frontline employees establish practices in accordance with specific rationales and to understand those rationales is the key to understand frontline employees. The identification of practices adds to the research on employment services and emphasizes the role of consultation as a key complementary to other measures including the allowances, agreements, brochures, or websites (Marston, 2006; Toerien et al., 2015; Boulus-Rødje, 2018). The rationales and motives that have been identified extends the line of research on the (implicit) drivers of consultants' behaviour in various advisory services (Dolata and Schwabe, 2017b; Dolata and Schwabe, 2018; Dolata et al., 2019). Finally, the study identifies the opportunities for design research oriented to support the consultants' mediator role rather than their singular activities.

## Acknowledgments

We want to express our gratitude to the branches of AA that were involved, as well as to the advisors who, in spite of initial concerns, provided the field researchers with valuable insights into their daily work. We also thank the many clients, who accepted the presence of the researchers during their consultations and who participated in the interviews. Our best thanks go to the anonymous reviewer team and to the program editors, whose comments motivated us to improve this manuscript.